\documentclass[12pt]{article}
\usepackage{hyperref}
\usepackage{amsmath}
\usepackage{graphicx}
\usepackage{amssymb} % Symbols for itemise
\usepackage{color}
\definecolor{mypurple}{rgb}{0.71,0.02,1}
\definecolor{myblue}{rgb}{0.14,0.11,0.49}
\newcommand{\Couleur}[1]{\textcolor{myblue}{#1}}
\newcommand{\Mat}[1]{{{\boldsymbol{#1}}}}
\newcommand{\abs}[1]{\left\vert#1\right\vert}
\def\be{\begin{equation}}
\def\ee{\end{equation}}
\def\bea{\begin{eqnarray}}
\def\eea{\end{eqnarray}}
\def\bc{\begin{center}}
\def\ec{\end{center}}
\def\bi{\begin{itemize}}
\def\ei{\end{itemize}}
\def\noi{\noindent}
\def\dd{\mathrm{d}}
\title{Summary of a non-uniqueness problem\\ of the covariant Dirac theory \\and of two solutions of it}
\author{
Mayeul Arminjon\\
\small\it Laboratory ``Soils, Solids, Structures, Risks'', 3SR \\
\small\it (CNRS \& Grenoble Universities: UJF, Grenoble-INP), Grenoble, France.
} % fin "author"
\date{}
\begin{document}
%%%%%%%%%%%%%%%%%%%%%%%%%%%%%%%%%%%%%%%%%%%%%%%%%%%%%%%%%%%%%%%%%%%%%%%%%%%%%%%%

\maketitle          % This won't show up when \onlynotestoo is in effect.

%%%%%%%%%%%%%%%%%%%%%%%%%%%%%%%%%%%%%%%%%%%%%%%%%%%%%%%%%%%%%%%%%%%%%%%%%%%%%%%%

%%%%%%%%%%%%%%%%%%%%%%%%%%%%%%%%%%%%%%%%%%%%%%%%%%%%%%%%%%%%%%%%%%%%%%%%%%%%%%%%

\begin{abstract} 
\noi We present a summary of: 1) the non-uniqueness problem of the Hamiltonian and energy operators associated, in any given coordinate system, with the generally-covariant Dirac equation; 2) two different ways to restrict the gauge freedom so as to solve that problem; 3) the application of these two ways to the case of a uniformly rotating reference frame in Minkowski spacetime: we find that a spin-rotation coupling term is there only with one of these two ways. 
\end{abstract}

%%%%%%%%%%%%%%%%%%%%%%%%%%%%%%%%%%%%%%%%%%%%%%%%%%%%%%%%%%%%%%%%%%%%%%%%%%%%%%%%
\section{Introduction}
%%%%%%%%%%%%%%%%%%%%%%%%%%%%%%%%%%%%%%%%%%%%%%%%%%%%%%%%%%%%%%%%%%%%%%%%%%%%%%%%
%%%%%%%%%%%%%%%%%%%%%%%%%%%%%%%%%%%%%%%%%%%%%%%%%%%%%%%%%%%%%%%%%%%%%%%%%%%%%%%%
\subsection{Experimental context}
%%%%%%%%%%%%%%%%%%%%%%%%%%%%%%%%%%%%%%%%%%%%%%%%%%%%%%%%%%%%%%%%%%%%%%%%%%%%%%%%

The following quantum effects in the classical gravitational field are observed on Earth for neutrons (which are spin $\frac{1}{2}$ particles) and for atoms:
\bi
\item The COW effect: the gravity-induced phase shift was measured by neutron \cite{COW1975} and atom \cite{RiehleBorde1991} interferometry;

\item The Sagnac effect: the Earth-rotation-induced phase shift was measured by neutron \cite{WernerStaudenmannColella1979} and atom \cite{KasevichChu1991} interferometry;

\item The Granit effect: the quantization of the energy levels was proved by observing a threshold in the neutron transmission through a thin horizontal slit \cite{Nesvizhevsky2002}.

\ei

\noi To this author's knowledge, these are the only observed effects of the gravity-quantum coupling. This motivates work on the curved-spacetime Dirac equation (thus first-quantized theory).

%%%%%%%%%%%%%%%%%%%%%%%%%%%%%%%%%%%%%%%%%%%%%%%%%%%%%%%%%%%%%%%%%%%%%%%%%%%%%%%%
  \subsection{State of the art}
%%%%%%%%%%%%%%%%%%%%%%%%%%%%%%%%%%%%%%%%%%%%%%%%%%%%%%%%%%%%%%%%%%%%%%%%%%%%%%%%

The  (generally-)covariant rewriting of the Dirac equation is \cite{BrillWheeler1957+Corr,ChapmanLeiter1976,deOliveiraTiomno1962}:
\be \label{Dirac-general}
\Couleur{\gamma ^\mu D_\mu\Psi=-iM\Psi \qquad (M\equiv mc/\hbar)}.
\ee
Here $\Couleur{\gamma ^\mu \ (\mu =0,...,3)}$ are the Dirac $4\times4$ matrices, which verify the anticommutation relation
\be\label{Clifford}
\Couleur{\gamma ^\mu \gamma ^\nu + \gamma ^\nu \gamma ^\mu = 2g^{\mu \nu}\,{\bf 1}_4, \ \mu ,\nu \in \{0,...,3\}, \ {\bf 1}_4\equiv \mathrm{diag}(1,1,1,1)},
\ee
where $\Couleur{(g^{\mu \nu})\equiv (g_{\mu \nu})^{-1}}$, with $\Couleur{g_{\mu \nu}}$ the components of the Lorentzian metric $\Couleur{\Mat{g}}$ on the spacetime manifold $\Couleur{\mathrm{V}}$ in a local chart $\Couleur{\chi : \mathrm{V}\supset \mathrm{U} \rightarrow  {\sf R}^4} $. \hypertarget{gamma = field}{Thus the $\Couleur{\gamma ^\mu }$ 's are defined locally and depend on $\Couleur{X \in \mathrm{U}}$}. In Eq. (\ref{Dirac-general}), $\Couleur{\Psi: \mathrm{U} \rightarrow  {\sf C}^4} $ is the local expression of the wave function $\Couleur{\psi }$ in a local frame field  $\Couleur{(e_a)_{a=0,...,3}}$ on $\Couleur{{\sf E }}$ over $\Couleur{\mathrm{U}}$, $\ \Couleur{\psi }$ itself being a section of the ``spinor bundle" $\Couleur{{\sf E }}$---i.e., a vector bundle with base $\Couleur{\mathrm{V}}$, such that, essentially, one can define a global ``\Couleur{$\gamma $} field" verifying an ``intrinsic" form of the relation (\ref{Clifford}) \cite{A45}. And $\Couleur{D_\mu \equiv \partial _\mu +\Gamma _\mu }$ are covariant derivatives, where $\Couleur{\Gamma _\mu }$ are the $4\times4$ connection matrices.\\

For the standard version of the covariant Dirac equation, which is due to Fock \& Weyl (hereafter the {\bf DFW} equation), the field of the Dirac matrices $\Couleur{\gamma ^\mu} $ is determined by an (orthonormal) {\it tetrad field} $\Couleur{(u_\alpha)}$, i.e., $\Couleur{u_\alpha}$ is a global vector field, for $\Couleur{\ \alpha =0,...,3}$. \hypertarget{local Lorentz}{The tetrad field $\Couleur{(u_\alpha)}$ may be changed} by a (smooth) ``local Lorentz transformation" $\Couleur{L: \mathrm{V} \rightarrow {\sf SO(1,3)}}$, $\ \Couleur{\widetilde{u}_\beta =L^\alpha _{\ \,\beta }u_\alpha}$. The latter can be always lifted to a smooth ``spin transformation" $\Couleur{S: \mathrm{V} \rightarrow {\sf Spin(1,3)}}$, provided that $\Couleur{\mathrm{V}}$ is topologically simple. Then the DFW equation is covariant under changes of the tetrad field, thus the DFW equation is unique \cite{BrillWheeler1957+Corr, ChapmanLeiter1976, deOliveiraTiomno1962}. \hypertarget{Spin depends on gamma}{That covariance is got with the ``spin connection"} $\Couleur{D}$ on the spinor bundle $\Couleur{{\sf E }}$. We note that the corresponding connection matrices $\Couleur{\Gamma _\mu }$ \cite{BrillWheeler1957+Corr, ChapmanLeiter1976, deOliveiraTiomno1962} depend explicitly on the tetrad field (thus on the field of the Dirac matrices $\Couleur{\gamma ^\mu} $), hence so does the ``spin connection" $\Couleur{D}$ itself. The DFW equation has been investigated in relevant physical situations, notably in rotating coordinates in a Minkowski spacetime (e.g. \cite{ChapmanLeiter1976, HehlNi1990}), in accelerating coordinates in a Minkowski spacetime (e.g. \cite{HehlNi1990, VarjuRyder1998, Obukhov2001}), in a static, or stationary, weak gravitational field (e.g. \cite{SoffelMullerGreiner1977, VarjuRyder1998, Obukhov2001, A38, Boulanger-Spindel2006}). The differences with the non-relativistic Schr\"odinger equation with Newtonian potential are not currently measurable. The first expected new effect with respect to that non-relativistic description is the ``spin-rotation coupling" in a rotating frame \cite{Mashhoon1988, HehlNi1990}. This effect would affect the energy levels of a Dirac particle.

%%%%%%%%%%%%%%%%%%%%%%%%%%%%%%%%%%%%%%%%%%%%%%%%%%%%%%%%%%%%%%%%%%%%%%%%%%%%%%%%
%%%%%%%%%%%%%%%%%%%%%%%%%%%%%%%%%%%%%%%%%%%%%%%%%%%%%%%%%%%%%%%%%%%%%%%%%%%%%%%%
  \subsection{Covariant Dirac equation: alternative versions}
%%%%%%%%%%%%%%%%%%%%%%%%%%%%%%%%%%%%%%%%%%%%%%%%%%%%%%%%%%%%%%%%%%%%%%%%%%%%%%%%

It was proved \cite{A45} that, for any physically relevant spacetime $\Couleur{\mathrm{V}}$, there are two explicit realizations of a spinor bundle $\Couleur{{\sf E}}$ : 
\bi
\item $\Couleur{{\sf E}=\mathrm{V}\times{\sf C}^4}$ (the wave function is a complex four-scalar)

\item $\Couleur{{\sf E}=\mathrm{T}_{\sf C}\mathrm{V}}\quad\,$ (the wave function is a complex {\it four-vector}).
\ei
\hypertarget{Alternative Eqs}{This motivates proposing} alternative versions of the covariant Dirac equation (\ref{Dirac-general}) \cite{A39,A45}, based on assuming any connection on either of these two spinor bundles---for example, the Levi-Civita connection, that is defined primarily on the real tangent bundle $\Couleur{\mathrm{TV}}$, but is straightforwardly extended to $\Couleur{\mathrm{T}_{\sf C}\mathrm{V}}$ \cite{A39}. Thus, the connection is fixed (\hyperlink{Spin depends on gamma}{in contrast with DFW}). The price to pay is that the covariance of the Dirac equation under changes of the $\Couleur{\gamma ^\mu} $ field, instead of being automatical as with DFW, is expressed by a system of quasilinear PDE's that depends on the starting $\Couleur{\gamma ^\mu} $ field \cite{A42,A43}. These alternative equations are actually more general than the DFW equation: given an arbitrary connection on either $\Couleur{\mathrm{V}\times{\sf C}^4}$ or $\Couleur{\mathrm{T}_{\sf C}\mathrm{V}}$, the DFW equation is equivalent to a particular case, obtained by choosing a particular $\Couleur{\gamma ^\mu} $ field, of the corresponding covariant Dirac equation \cite{A45}.

%%%%%%%%%%%%%%%%%%%%%%%%%%%%%%%%%%%%%%%%%%%%%%%%%%%%%%%%%%%%%%%%%%%%%%%%%%%%%%%%
  \subsection{Surprising recent results}
%%%%%%%%%%%%%%%%%%%%%%%%%%%%%%%%%%%%%%%%%%%%%%%%%%%%%%%%%%%%%%%%%%%%%%%%%%%%%%%%

Ryder \cite{Ryder2008} considered uniform rotation with respect to an inertial frame in a Minkowski spacetime. He found that, in this particular case, Mashhoon's term in the DFW Hamiltonian operator $\Couleur{\mathrm{H}}$ is there for one tetrad field $\Couleur{(u_\alpha )}$, but is not there for another one, say $\Couleur{(\widetilde{u}_\alpha )}$. Independently, in the most general case, the relevant scalar product for the covariant Dirac equation was identified, and it was found that the hermiticity of $\Couleur{\mathrm{H}}$ w.r.t. that scalar product depends on the choice of the admissible field $\Couleur{\gamma ^\mu }$ \cite{A42}. This fact (the instability of the hermiticity of $\Couleur{\mathrm{H}}$ under the admissible changes of the $\Couleur{\gamma ^\mu }$ field) meant there is a non-uniqueness problem in the covariant Dirac theory and asked for a general study of this problem. As for this fact, that study was done for DFW, {\it and} for \hyperlink{Alternative Eqs}{alternative versions} of the covariant Dirac equation. It was found \cite{A43} that, for the standard version as well as the alternative ones, in any given reference frame: i) The Hamiltonian operator $\Couleur{\mathrm{H}}$ is non-unique. (Concrete examples of this non-uniqueness have been shown in Ref. \cite{GorbatenkoNeznamov2010}.) ii) So is also the energy operator $\Couleur{\mathrm{E}}$ (which coincides with the Hermitian part of $\Couleur{\mathrm{H}}$ \cite{Leclerc2006,A43}). iii) The Dirac energy spectrum (i.e., the spectrum of $\Couleur{\mathrm{E}}$) is non-unique. Let us briefly review this problem.

%%%%%%%%%%%%%%%%%%%%%%%%%%%%%%%%%%%%%%%%%%%%%%%%%%%%%%%%%%%%%%%%%%%%%%%%%%%
\section{Summary of the non-uniqueness problem}

%%%%%%%%%%%%%%%%%%%%%%%%%%%%%%%%%%%%%%%%%%%%%%%%%%%%%%%%%%%%%%%%%%%%%%%%%%%%%%%%
   
%%%%%%%%%%%%%%%%%%%%%%%%%%%%%%%%%%%%%%%%%%%%%%%%%%%%%%%%%%%%%%%%%%%%%%%%%%%%%%%%

\subsection{Local similarity (or gauge) transformations}

If one changes from one field of Dirac matrices \Couleur{$(\gamma ^\mu)$} to another one \Couleur{$(\widetilde {\gamma }^\mu)$}, also satisfying the anticommutation relation (\ref{Clifford}), the new field obtains by a {\it local similarity transformation} (or local gauge transformation): for any \Couleur{$X \in \mathrm{V}$} there is an invertible complex $4\times 4$ matrix \Couleur{$S(X) $}, such that [for any chart \Couleur{$X\mapsto (x^\mu )$}, and for any \Couleur{$X $} in the domain of that chart]: 
\be \label{similarity-gamma}
\Couleur{\widetilde{\gamma} ^\mu(X)  =  S(X)^{-1}\gamma ^\mu(X) S(X), \quad \mu =0,...,3}.
\ee

\vspace{2mm}
\noi For the {\it standard} covariant Dirac equation (DFW), {\it the admissible local gauge transformations are the (smooth) mappings} \Couleur{$\mathrm{V}\rightarrow  {\sf Spin(1,3)}$}, because they are got by \hyperlink{local Lorentz}{lifting a (smooth) local Lorentz transformation} \Couleur{$L(X)$} applied to a tetrad field. Only for the {\it alternative} versions briefly presented \hyperlink{Alternative Eqs}{above}, the local gauge transformations are more general: \Couleur{$S(X) \in {\sf GL}(4,{\sf C})$}---but then also, as mentioned there, only a subgroup of the group of the smooth gauge transformations \Couleur{$\mathrm{V}\rightarrow {\sf GL}(4,{\sf C})$} leaves the Dirac equation covariant. The existence of the non-uniqueness problem reviewed below has been proved in greater detail for DFW and, while doing this, it was explicitly accounted for its admissible local gauge transformations \Couleur{$S(X) \in {\sf Spin(1,3)}$} \cite{A43}. 
   
%%%%%%%%%%%%%%%%%%%%%%%%%%%%%%%%%%%%%%%%%%%%%%%%%%%%%%%%%%%%%%%%%%%%%%%%%%%%%%%%

\subsection{The general Dirac Hamiltonian}

Rewriting the covariant Dirac equation in the ``Schr\"odinger" form:
\be \label{Schrodinger-general}
\Couleur{i \frac{\partial \Psi }{\partial t}= \mathrm{H}\Psi,\qquad (t\equiv x^0)},
\ee
gives the general explicit expression of the Hamiltonian operator \Couleur{$\mathrm{H}$} \cite{A38,A43}. An important point is that \hypertarget{reference-frame}{\Couleur{$\mathrm{H}$} depends on the coordinate system}, or more exactly on the {\it reference frame} \cite{A42}---defined formally \cite{A44} as an equivalence class of charts defined on a given open set $\Couleur{\mathrm{U}\subset \mathrm{V}}$ and exchanging by 
\be\label{purely-spatial-change}
\Couleur{x'^0=x^0,\quad x'^j=f^j((x^k)) \qquad (j,k=1,2,3)}.
\ee
(Note that a chart $\Couleur{\chi }$ defines thus a reference frame: the equivalence class of $\Couleur{\chi }$.) The dependence of the Hamiltonian operator \Couleur{$\mathrm{H}$} on the reference frame is valid for any wave equation and has nothing to do with the non-uniqueness problem reviewed here, which problem applies to the covariant Dirac equation with its gauge freedom.

%%%%%%%%%%%%%%%%%%%%%%%%%%%%%%%%%%%%%%%%%%%%%%%%%%%%%%%%%%%%%%%%%%%%%%%%%%%%%%%%

%%%%%%%%%%%%%%%%%%%%%%%%%%%%%%%%%%%%%%%%%%%%%%%%%%%%%%%%%%%%%%%%%%%%%%%%%%%%%%%%
   
%%%%%%%%%%%%%%%%%%%%%%%%%%%%%%%%%%%%%%%%%%%%%%%%%%%%%%%%%%%%%%%%%%%%%%%%%%%%%%%%
\subsection{Invariance condition of the Hamiltonian under a local gauge transformation}

Consider some wave equation and apply a local gauge transformation \Couleur{$S$} to the field of its coefficients (here the field of Dirac matrices \Couleur{$\gamma ^\mu$}). We assume that \Couleur{$S$} allows one to define an isometry between the two Hilbert spaces before and after the transformation, as is the case for the covariant Dirac equation \cite{A43}. The necessary and sufficient condition in order that the Hamiltonians before and after the gauge transformation \Couleur{$S$}, \Couleur{$\mathrm{H}$} and \Couleur{$\widetilde{\mathrm{H}}$}, be physically equivalent [i.e., the condition in order that all scalar products of the form \Couleur{$(\Phi \mid \mathrm{H}\Psi )$} be invariant], is \cite{A43}:
\be \label{similarity-invariance-H}
\Couleur{\widetilde{\mathrm{H} }  =  S^{-1}\,\mathrm{H}\, S}.
\ee
Let us ask when this condition is fulfilled. E.g. if the wave equation is covariant under \Couleur{$S$}, it is easy to see \cite{A47} that \hypertarget{H invariance DFW}{we have (\ref{similarity-invariance-H}) iff \Couleur{$S(X)$} is time-independent}, 
\be\label{d_0 S=0}
\Couleur{\partial_0 S=0},
\ee
independently of the explicit form of \Couleur{$\mathrm{H}$}, thus independently of the wave equation. Now, as mentioned above, the DFW equation is indeed covariant when \Couleur{$S$} is an admissible gauge transformation for DFW, i.e., when it \hyperlink{local Lorentz}{takes values in ${\sf Spin(1,3)}$}. So, for DFW, Eq. (\ref{d_0 S=0}) is really the condition in order that an admissible gauge transformation lead to an equivalent Hamiltonian. \{For the alternative equations, the condition to have (\ref{similarity-invariance-H}) is a bit less simple \cite{A43}.\}\\

However, in the general case: \Couleur{$g_{\mu \nu ,0} \ne 0$}, any possible field \Couleur{$\gamma ^\mu$} depends on \Couleur{$\ t\ $}, and so does generally \Couleur{$S$}, thus the condition (\ref{d_0 S=0}) is not verified. Thus {\it the Dirac Hamiltonian is not unique} (even in a given coordinate system) and one also proves that {\it the energy operator and its spectrum are not unique}  \cite{A43}. The physical relevance of the energy operator is justified by two facts: i) it is the Hermitian part of the Hamiltonian operator, and ii) its mean value is the field energy. See App. B of Ref. \cite{A48}.

%%%%%%%%%%%%%%%%%%%%%%%%%%%%%%%%%%%%%%%%%%%%%%%%%%%%%%%%%%%%%%%%%%%%%%%%%%%%%%%%

%%%%%%%%%%%%%%%%%%%%%%%%%%%%%%%%%%%%%%%%%%%%%%%%%%%%%%%%%%%%%%%%%%%%%%%%%%%%%%%%
  \subsection{Basic reason for the non-uniqueness}
%%%%%%%%%%%%%%%%%%%%%%%%%%%%%%%%%%%%%%%%%%%%%%%%%%%%%%%%%%%%%%%%%%%%%%%%%%%%%%%%
%Thus, in a given general reference frame or even in a given coordinate system, the Hamiltonian and energy operators associated with the {\it generally-covariant} Dirac equation depend on the choice of the {\it field} of Dirac matrices $\Couleur{X \mapsto \gamma ^\mu(X)}$. In contrast, 
The Hamiltonian operator associated, in any given Cartesian coordinate system, with the {\it original} Dirac equation of special relativity is Hermitian and does not depend on the choice of the {\it constant} set of Dirac matrices $\Couleur{\gamma ^{\sharp \alpha }}$ \cite{A40}. Thus we have only constant gauge transformations for the original Dirac theory and as a result the non-uniqueness problem is absent from that theory. The DFW theory, on the other hand, has been built so that the DFW {\it equation} be covariant under the smooth ${\sf Spin(1,3)}$ transformations. Yet it turns out that the associated {\it energy operator} is {\it not} invariant under these gauge transformations. Now the principle that ``physical observables are gauge invariant" cannot disqualify the energy operator, because this is the most important quantum-mechanical observable. Thus, what this principle tells us in that instance is that we have to restrict the gauge freedom.

%%%%%%%%%%%%%%%%%%%%%%%%%%%%%%%%%%%%%%%%%%%%%%%%%%%%%%%%%%%%%%%%%%%%%%%%%%%%%%%%
\section{A ``conservative" solution of the non-uniqueness problem}
%%%%%%%%%%%%%%%%%%%%%%%%%%%%%%%%%%%%%%%%%%%%%%%%%%%%%%%%%%%%%%%%%%%%%%%%%%%%%%%%

%%%%%%%%%%%%%%%%%%%%%%%%%%%%%%%%%%%%%%%%%%%%%%%%%%%%%%%%%%%%%%%%%%%%%%%%%%%%%%%%
  \subsection{Tetrad fields adapted to a reference frame}
%%%%%%%%%%%%%%%%%%%%%%%%%%%%%%%%%%%%%%%%%%%%%%%%%%%%%%%%%%%%%%%%%%%%%%%%%%%%%%%%

The data of a physically admissible \hyperlink{reference-frame}{reference frame} $\Couleur{\mathrm{F}}$ fixes a unique four-velocity field $\Couleur{v_\mathrm{F}}$: the unit tangent vector to the world lines 
\be\label{x^j = constant}
\Couleur{X \in \mathrm{U}, \quad x^0(X) \mathrm{\ variable,\qquad }x^j(X) =\mathrm{constant\ for\ }j=1,2,3}.
\ee
(The ``physical admissibility" means precisely that these world lines are time-like, which is true iff $g_{00}>0$ \cite{A44}.) These world lines, which are invariant under an internal change (\ref{purely-spatial-change}), are the trajectories of the particles constituting the reference frame \cite{A44,Cattaneo1958}. Thus, a physically admissible chart {\it has} physical content after all. It is natural to impose on the tetrad field $\Couleur{(u_\alpha )}$ the condition that the time-like vector of the tetrad be the four-velocity of the reference frame:
\be\label{u_0=v_F}
\Couleur{u_0=v_\mathrm{F}}.
\ee
Then the spatial triad $\Couleur{(u_p) \quad (p=1,2,3)}$ can only be {\it rotating} w.r.t. the reference frame \cite{A47}. (An outline follows.)%This is defined by a spatial rotation rate tensor field $\Couleur{\Mat{\Xi}}$. 

%%%%%%%%%%%%%%%%%%%%%%%%%%%%%%%%%%%%%%%%%%%%%%%%%%%%%%%%%%%%%%%%%%%%%%%%%%%%%%%%

%%%%%%%%%%%%%%%%%%%%%%%%%%%%%%%%%%%%%%%%%%%%%%%%%%%%%%%%%%%%%%%%%%%%%%%%%%%%%%%%
  \subsection{Space manifold and spatial tensor fields}
%%%%%%%%%%%%%%%%%%%%%%%%%%%%%%%%%%%%%%%%%%%%%%%%%%%%%%%%%%%%%%%%%%%%%%%%%%%%%%%%

Let $\Couleur{\mathrm{F}}$ be a reference frame, with its domain $\Couleur{\mathrm{U}\subset \mathrm{V}}$. The set $\Couleur{\mathrm{M}}$ of the world lines (\ref{x^j = constant}) is endowed with a natural structure of differential manifold: for any chart $\Couleur{\chi \in \mathrm{F}}$, its spatial part allows us to define a mapping $\Couleur{\widetilde{\chi}: \mathrm{M} \ni x \mapsto (x^j)_{j=1,2,3}}$, which is a chart on $\Couleur{\mathrm{M}}$. Thus, the space manifold $\Couleur{\mathrm{M}}$ is frame-dependent and is {\it not} a 3-D submanifold of the spacetime manifold $\Couleur{ \mathrm{V}}$ \cite{A44}. One then defines \cite{A47} spatial tensor fields depending on the spacetime position, e.g. a spatial vector field: $\Couleur{\mathrm{U} \ni X \mapsto {\bf u}(X) \in \mathrm{TM}_{x(X)}}$, where, for $\Couleur{X \in \mathrm{U}}$, $\Couleur{x(X)}$ is the unique world line $\Couleur{x \in \mathrm{M}}$, such that $\Couleur{X \in x}$. [See Eq. (\ref{x^j = constant}).]

%%%%%%%%%%%%%%%%%%%%%%%%%%%%%%%%%%%%%%%%%%%%%%%%%%%%%%%%%%%%%%%%%%%%%%%%%%%%%%%%

%%%%%%%%%%%%%%%%%%%%%%%%%%%%%%%%%%%%%%%%%%%%%%%%%%%%%%%%%%%%%%%%%%%%%%%%%%%%%%%%
  \subsection{Rotation rate tensor field of the spatial triad}
%%%%%%%%%%%%%%%%%%%%%%%%%%%%%%%%%%%%%%%%%%%%%%%%%%%%%%%%%%%%%%%%%%%%%%%%%%%%%%%%

Again a reference frame $\Couleur{\mathrm{F}}$ is given. For any $\Couleur{ X \in \mathrm{U}}$, there is a canonical isomorphism $i_X$ between, on one hand, the hyperplane \Couleur{$\mathrm{H}_X$} of the four-vectors which are orthogonal to $\Couleur{v_\mathrm{F}}$ [the unit tangent vector to the world lines (\ref{x^j = constant})] and, on the other hand, the vector space \Couleur{$\mathrm{TM}_{x(X)}$} of the spatial vectors at \Couleur{$x(X)$}:
\be\label{H_X}
 \Couleur{\mathrm{H}_X \equiv \{ u_X \in \mathrm{TV}_X\,;\ \Mat{g}(u_X,v_\mathrm{F}(X))=0 \} \rightleftharpoons \mathrm{TM}_{x(X)}}.
\ee
To the spacetime vector $\Couleur{u \in \mathrm{H}_X}$, with components $\ \Couleur{u^\mu\ (\mu =0,...,3)}$ in some chart $\Couleur{\chi \in \mathrm{F}}$, the isomorphism \Couleur{$i_X$} associates the spatial vector $\Couleur{{\bf u } \in \mathrm{TM}_{x(X)}}$, whose components are simply $\Couleur{u^j\ (j =1,2,3})$ in the associated chart $\Couleur{\widetilde{\chi }}$. This is independent of the chart $\Couleur{\chi \in \mathrm{F}}$ \cite{A47}.\\

Then, there is one natural time-derivative for spatial vectors: the Fermi-Walker derivative applied to a spatial vector $\Couleur{{\bf u}(\xi) }$, \Couleur{$\delta {\bf u}/d\xi $}, which is relative to a given four-velocity field, here \Couleur{$v_\mathrm{F}$} \cite{A47}. This allows us to define the rotation rate tensor $\Couleur{\Mat{\Xi}(\xi )}$, along a curve \Couleur{$C:\ \xi \mapsto X(\xi )$} in the spacetime, of the spatial triad field $\Couleur{({\bf u}_p) \ \, (p=1,2,3)}$ associated with a tetrad field $\Couleur{(u_\alpha ) \ \, (\alpha =0,...,3)}$: $\Couleur{\Mat{\Xi}}$ is such that 
\be\label{Xi _pq}
\Couleur{\Xi _{p q } = \Mat{h}\left({\bf u}_p ,\left(\frac{\delta {\bf u}_q }{d\xi }\right) \right)}=-\Couleur{\Xi _{q p }}\quad (\Couleur{p,q=1,2,3}),
\ee
where $\Couleur{\Mat{h}}$ is the spatial metric in the reference frame $\Couleur{\mathrm{F}}$. At any \Couleur{$X \in \mathrm{U}$}, we take the world line \Couleur{$x(X)$}, parameterized by the coordinate time \Couleur{$t$}, as the curve \Couleur{$C$}. We get thus the rotation rate tensor field $\Couleur{\Mat{\Xi}(X )}$ of the spatial triad, along the world lines (\ref{x^j = constant}). We have explicitly \cite{A47}:
\be\label{Xi explicit}
\Couleur{\Xi _{p q}=-c\frac{\dd \tau }{\dd t}\gamma _{pq0}},
\ee
where \Couleur{$\gamma _{\alpha \beta \epsilon }\equiv \eta _{\alpha \zeta}\gamma^\zeta_{\ \beta \epsilon } $} and the \Couleur{$\gamma^\zeta_{\ \beta \epsilon }$} 's are the coefficients of the Levi-Civita connection, when an orthonormal tetrad field $(u_\alpha)$ is taken as the frame field \cite{A47,Ryder2008}. ($\tau $ is the proper time along $x(X)$ and $\eta $ is the Minkowski metric.) We prove then that two tetrad fields $\Couleur{(u_\alpha )}$ and $\Couleur{(\widetilde{u}_\alpha )}$ such that $\Couleur{u_0=\widetilde{u}_0=v_\mathrm{F}}$, and which have the same rotation rate field $\Couleur{\Mat{\Xi}=\widetilde{\Mat{\Xi}}}$, exchange by a time-independent Lorentz transformation. Hence they give rise \underline{in $\Couleur{\mathrm{F}}$} to equivalent Hamiltonian operators and to equivalent energy operators  \cite{A47}. \\

Two natural ways to fix the tensor field $\Couleur{\Mat{\Xi}}$ are: i) $\Couleur{\Mat{\Xi}=\Mat{\Omega }}$, where $\Couleur{\Mat{\Omega }}$ is the unique {\it rotation rate field of the given reference frame} $\Couleur{\mathrm{F}}$ \cite{A47,Cattaneo1958}, and ii) $\Couleur{\Mat{\Xi}=\Mat{0}}$. Either choice, i) or ii), thus provides a solution to the non-uniqueness problem. These two solutions are not equivalent, so that experiments would be required to decide between the two. Moreover, each solution is valid {\bf only in a given reference frame}.

%%%%%%%%%%%%%%%%%%%%%%%%%%%%%%%%%%%%%%%%%%%%%%%%%%%%%%%%%%%%%%%%%%%%%%%%%%%%%%%%

%%%%%%%%%%%%%%%%%%%%%%%%%%%%%%%%%%%%%%%%%%%%%%%%%%%%%%%%%%%%%%%%%%%%%%%%%%%%%%%%
 \section{Getting unique Hamiltonian and energy operators in any reference frame at once?}

%%%%%%%%%%%%%%%%%%%%%%%%%%%%%%%%%%%%%%%%%%%%%%%%%%%%%%%%%%%%%%%%%%%%%%%%%%%%%%%%

The invariance condition of the Hamiltonian \Couleur{$\mathrm{H}$} after a local gauge transformation for DFW: \hyperlink{H invariance DFW}{\Couleur{$\partial_0 S=0$}}, is coordinate-dependent. This condition implies also the invariance of the energy operator \Couleur{$\mathrm{E}$} for DFW \cite{A43}. Therefore, the stronger condition \Couleur{$\partial_\mu S=0\ (\mu =0,...,3)$} implies the invariance of both \Couleur{$\mathrm{H}$} and \Couleur{$\mathrm{E}$} simultaneously in any chart (hence in any reference frame), for DFW.\\

On the other hand, for the alternative versions of the covariant Dirac equation, the invariance conditions of \Couleur{$\mathrm{H}$} and \Couleur{$\mathrm{E}$} are somewhat less simple: they contain the covariant derivatives \Couleur{$D_\mu S$} \cite{A43}. But, \hypertarget{QRD-0}{for the ``QRD--0" version}, we define the connection matrices to be \cite{A45}:
\be
\Couleur{\Gamma _\mu =0} \quad \mathrm{in\ the\ canonical\ frame\ field\ } \Couleur{(E_a)}\ \mathrm{of}\ \Couleur{\mathrm{V}\times{\sf C}^4},
\ee
so we have by construction \Couleur{$\partial_\mu S=D_\mu S$} for QRD--0.\\

\hypertarget{DFW+QRD0}{Thus, if we succeed in restricting} the choice of the $\Couleur{\gamma ^\mu }$ field so that any two choices exchange by a {\bf constant} gauge transformation (\Couleur{$\partial_\mu S=0$}), then we solve the non-uniqueness problem simultaneously in any reference frame---for both DFW and QRD--0, and only for them.

%%%%%%%%%%%%%%%%%%%%%%%%%%%%%%%%%%%%%%%%%%%%%%%%%%%%%%%%%%%%%%%%%%%%%%%%%%%%%%%%

%%%%%%%%%%%%%%%%%%%%%%%%%%%%%%%%%%%%%%%%%%%%%%%%%%%%%%%%%%%%%%%%%%%%%%%%%%%%%%%%
  \subsection{Fixing one tetrad field in each chart}
%%%%%%%%%%%%%%%%%%%%%%%%%%%%%%%%%%%%%%%%%%%%%%%%%%%%%%%%%%%%%%%%%%%%%%%%%%%%%%%%

In a chart, a tetrad $\Couleur{(u_\alpha)}$ is defined by a $4\times 4$ real matrix $\Couleur{a\equiv (a^\mu  _{\ \,\alpha})}$, such that $\Couleur{u_\alpha = a^\mu _{\ \, \alpha } \partial _\mu }$. \hypertarget{Orthonormal b}{The orthonormality condition for a tetrad} in the metric with matrix $\Couleur{G\equiv (g_{\mu \nu })}$ \Couleur{$=G(X) \quad (X \in \mathrm{V})$} writes:
\be\label{Lorentz-Cholesky}
\Couleur{b^T\eta b = G \qquad [b\equiv a^{-1},\quad \eta \equiv \mathrm{diag}(1,-1,-1,-1)]}.
\ee

\vspace{2mm}
\noi The classical Cholesky decomposition, which applies to a positive-definite symmetric matrix, can be extended to the matrix of a Lorentzian metric \cite{A48,Reifler2008}: There is a unique lower triangular solution $\ \Couleur{b=C}$ of (\ref{Lorentz-Cholesky}) with $\Couleur{C^{\mu }_{\ \,\mu }>0,\ \mu =0,...,3}$. This decomposition thus provides a unique tetrad in a given chart. Call this the ``Cholesky prescription". We know of one other prescription with this property \cite{Kibble1963}. Both coincide for a ``diagonal metric": for both, if $\Couleur{G = \mathrm{diag}(d_\mu ) }$, we get 
\be\label{diag tetrad}
\Couleur{\ u_\alpha  \equiv \delta _\alpha ^\mu \,\partial _\mu /\sqrt{\abs{d_\mu }}}. 
\ee
This is the ``diagonal tetrad" prescription.

\subsection{Fixing one tetrad field in each chart is not enough}

 What is physically given is the reference frame: a three-dimensional congruence of time-like world lines. Given a reference frame $\Couleur{\mathrm{F}}$, there remains a whole functional space of different choices for a chart $\Couleur{\chi \in \mathrm{F}}$, Eq. (\ref{purely-spatial-change}). Consider a prescription (e.g. ``Cholesky"): $\Couleur{\chi \mapsto a \mapsto (u_\alpha )}$. For two different charts $\Couleur{\chi,\chi '\in \mathrm{F}}$, we get two tetrad fields $\Couleur{(u_\alpha ),\ (u'_\alpha )}$ with matrices $\Couleur{a,a'}$. We have $\Couleur{u'_\beta  =L^\alpha _{\ \,\beta }u_\alpha}$, with
\be\label{L for two tetrads}
\Couleur{L=b\,P\,a', \quad b\equiv a^{-1},\qquad P^\mu _{\ \,\nu  }\equiv \frac{\partial x^\mu }{\partial x'^\nu }}.
\ee

\vspace{1mm}
Since the matrices $\Couleur{G}$ and $\Couleur{G'}$ depend on $\Couleur{t\equiv x^0=x'^0}$, so do $\Couleur{b}$ and $\Couleur{a'}$, Eq. (\ref{Lorentz-Cholesky}). Since  $\Couleur{\chi,\chi '\in \mathrm{F}}$, the matrix $\Couleur{P}$ does not depend on $\Couleur{t}$, Eq. (\ref{purely-spatial-change}). In general, the dependences on $\Couleur{t}$ of $\Couleur{b}$ and $\Couleur{a'}$ do not cancel each other in Eq. (\ref{L for two tetrads}). Thus in general the Lorentz transformation $\Couleur{L}$ depends on $\Couleur{t}$. Therefore, \hyperlink{Spin depends on gamma}{$\Couleur{L}$ is lifted to a gauge transformation $\Couleur{S}$} depending on $\Couleur{t}$.  According to Eq. (\ref{d_0 S=0}), it follows that
$\Couleur{\mathrm{H}}$ and $\Couleur{\mathrm{H}'}$ are not equivalent: {\it The non-uniqueness is still there} \cite{A48}.

%%%%%%%%%%%%%%%%%%%%%%%%%%%%%%%%%%%%%%%%%%%%%%%%%%%%%%%%%%%%%%%%%%%%%%%%%%%%%%%%
  \subsection{The case with a diagonal metric}
%%%%%%%%%%%%%%%%%%%%%%%%%%%%%%%%%%%%%%%%%%%%%%%%%%%%%%%%%%%%%%%%%%%%%%%%%%%%
Consider the Cholesky prescription applied to a ``diagonal metric": $\Couleur{G = \mathrm{diag}(d_\mu )}$, with \Couleur{$d_0>0,\quad d_j<0, j=1,2,3$}. Some algebra gives us \cite{A48}
\be\label{L^p_3-4}
\Couleur{\frac{\partial }{\partial t}\left(L^p  _{\ \,3 }\right)  \propto P^p   _{\ \,3} (P^j   _{\ \,3})^2 \frac{\partial }{\partial t}\left(\frac{d_j}{d_p}\right) \quad (\mathrm{no\ sum\ on\ }p=1,2,3)},
\ee
with a non-zero proportionality factor. Thus in general $\Couleur{\frac{\partial }{\partial t}\left(L^p  _{\ \,3 }\right) \ne 0}$, so again the non-uniqueness of $\Couleur{\mathrm{H}}$ and $\Couleur{\mathrm{E}}$ is still there.\\

An exception is when $\Couleur{d_j(X)=d_j^0\, h(X)}$ with $\Couleur{d_j^0}$ constant ($\Couleur{d_j^0 < 0}$ with $\Couleur{h > 0}$). Then, after changing $\Couleur{x'^j=x^j\,\sqrt{ -d_j^0}}$, we get $\Couleur{d'_j=-h \ (j=1,2,3)}$, or
\be\label{space-isotropic diagonal}
\Couleur{G\equiv (g_{\mu \nu })=\mathrm{diag}(f,-h,-h,-h)}, \qquad \Couleur{f>0,\ h>0}.
\ee

\vspace{2mm}
\noi That case provides us with a solution of the non-uniqueness problem that applies simultaneously in any reference frame:\\

\noi {\bf Theorem} \cite{A48}. {\it Let the metric have the space-isotropic diagonal form (\ref{space-isotropic diagonal}) in some chart $\Couleur{\chi }$. Let $\Couleur{\chi'}$ belong to the same reference frame $\Couleur{\mathrm{R}}$.} \\

(i) {\it The metric has the} {\bf form} {\it (\ref{space-isotropic diagonal}) also in $\Couleur{\chi'}$, iff  $\Couleur{(x^j)\mapsto (x'^j)}$ is a constant rotation, combined with a constant homothecy.}\\

(ii) {\it If one applies the ``diagonal tetrad" prescription (\ref{diag tetrad}) in each of the two charts, the two tetrads obtained thus are related together by a} {\bf constant} {\it Lorentz transformation $\Couleur{L}$, hence give rise,} {\bf in any reference frame} $\Couleur{\mathrm{F}}$, {\it to equivalent Hamiltonian operators as well to equivalent energy operators---\hyperlink{DFW+QRD0}{for the DFW and QRD--0 versions} of the Dirac equation.}

%%%%%%%%%%%%%%%%%%%%%%%%%%%%%%%%%%%%%%%%%%%%%%%%%%%%%%%%%%%%%%%%%%%%%%%%%%%%%%%%

%%%%%%%%%%%%%%%%%%%%%%%%%%%%%%%%%%%%%%%%%%%%%%%%%%%%%%%%%%%%%%%%%%%%%%%%%%%%%%%%
  \section{Application: uniformly rotating frame in flat spacetime}
%%%%%%%%%%%%%%%%%%%%%%%%%%%%%%%%%%%%%%%%%%%%%%%%%%%%%%%%%%%%%%%%%%%%%%%%%%%%

Let $\Couleur{\chi ': X\mapsto (ct',x',y',z')}$ be a Cartesian chart in a Minkowski spacetime, thus $\Couleur{g'_{\mu \nu }=\eta _{\mu \nu }}$. This defines an inertial reference frame $\Couleur{\mathrm{F}'}$. Then go from $\Couleur{\chi '}$ to $\Couleur{\chi: X\mapsto (ct,x,y,z)}$ defining a uniformly rotating reference frame $\Couleur{\mathrm{F}}$ ($\Couleur{\omega =\mathrm{constant}}$): 
\be\label{rotating Cartesian-1}
\Couleur{t=t',\ x=x'\cos \omega t + y' \sin \omega t,\ y=-x' \sin \omega t + y' \cos \omega t,\ z=z'}.
\ee
With $\Couleur{\rho \equiv \sqrt{x^2+y^2}}$, the Minkowski metric writes in the chart $\Couleur{\chi}$ (e.g. \cite{A47}):
\be\label{Minkowski in rotating Cartesian}%\nonumber
\Couleur{g_{00}=1-\left(\frac{\omega \rho }{c}\right)^2,\quad g_{01}=\frac{\omega y}{c}, \quad g_{02}=-\frac{\omega x}{c},\quad g_{03}=0,\quad g_{jk}=-\delta _{jk}}.
\ee
The 4-velocity of $\Couleur{\mathrm{F}}$ is $\Couleur{\ v=\partial _0/\sqrt{g_{00}}}$ \cite{A47}. Therefore, $\Couleur{\Mat{g}(v,\partial _j) \ne 0 \ (j=1,2)}$, where $\Couleur{(\partial _\mu )}$ is the natural basis of the ``rotating chart" $\Couleur{\chi}$. One may note \cite{A47} that each of Ryder's \cite{Ryder2008} two tetrads has $\Couleur{u_0=v'}$ (that is, the four-velocity of the inertial frame $\Couleur{\mathrm{F}'}$), hence $\Couleur{u_0\ne v}$. In this sense, each of Ryder's two tetrads is adapted to the inertial frame, not to the rotating frame.

%%%%%%%%%%%%%%%%%%%%%%%%%%%%%%%%%%%%%%%%%%%%%%%%%%%%%%%%%%%%%%%%%%%%%%%%%%%%%%%%

%%%%%%%%%%%%%%%%%%%%%%%%%%%%%%%%%%%%%%%%%%%%%%%%%%%%%%%%%%%%%%%%%%%%%%%%%%%%%%%%
  \subsection{A tetrad adapted to the rotating frame}
%%%%%%%%%%%%%%%%%%%%%%%%%%%%%%%%%%%%%%%%%%%%%%%%%%%%%%%%%%%%%%%%%%%%%%%%%%%%%%%%{\bf 

From now on, we are announcing results that will be presented in more detail elsewhere. Adopt the ``rotating cylindrical" chart $\Couleur{\chi ^\circ }$, also belonging to the rotating frame $\Couleur{\mathrm{F}}$. It is related to the ``rotating Cartesian" chart (\ref{rotating Cartesian-1}) by
\be\label{rotating Cartesian}
\Couleur{\chi ^\circ : X\mapsto (ct,\rho, \varphi , z)} \quad \mathrm{with}\ \ \Couleur{x=\rho \cos\varphi,\quad y=\rho  \sin\varphi }.
\ee
Define the following tetrad:
\be\label{rotating tetrad}
\Couleur{u_0\equiv v, \ u_p\equiv \Pi \partial _p/\parallel \Pi \partial _p\parallel},
\ee
where $\Couleur{\Pi=\Pi _X}$ is the orthogonal projection onto the hyperplane $\Couleur{\mathrm{H}_X}$ that is orthogonal to $\Couleur{v(X)}$. This is an {\it orthonormal} tetrad adapted to $\Couleur{\mathrm{F}}$, because for the chart $\Couleur{\chi ^\circ }$ we have $\Couleur{\Mat{g}(u_p,u_q)=0}$ for $\Couleur{\ 1\leq \ p \ne q \leq 3}$. The rotation rate tensor of $\Couleur{({\bf u}_p )}$ is given by Eq. (\ref{Xi explicit}). Here we find $\ \Couleur{\Xi _{p q}=0}$, except for 
\be
\Couleur{\Xi _{21}=-\Xi _{12}=\frac{\omega }{\sqrt{1-(\omega \rho )^2/c^2 }}}.
\ee
This differs from the rotation rate tensor $\Couleur{\Mat{\Omega }}$ of the {\it rotating frame} $\Couleur{\mathrm{F}}$ \cite{A47} only by $\Couleur{O(V^2/c^2)}$ terms (for $\Couleur{V\equiv \omega \rho \ll c}$).

%%%%%%%%%%%%%%%%%%%%%%%%%%%%%%%%%%%%%%%%%%%%%%%%%%%%%%%%%%%%%%%%%%%%%%%%%%%%%%%%

%%%%%%%%%%%%%%%%%%%%%%%%%%%%%%%%%%%%%%%%%%%%%%%%%%%%%%%%%%%%%%%%%%%%%%%%%%%%%%%%
  \subsection{Hamiltonian operator in the rotating frame with two different tetrads}
%%%%%%%%%%%%%%%%%%%%%%%%%%%%%%%%%%%%%%%%%%%%%%%%%%%%%%%%%%%%%%%%%%%%%%%%%%%%%%%%{\bf 

The Hamiltonian operator for the generally-covariant Dirac equation (\ref{Dirac-general}) is \cite{A42}:
\be \label{Hamilton-Dirac-normal}
\Couleur{\mathrm{H} =  mc^2\alpha  ^0 -i\hbar c\alpha ^j D _j -i\hbar c\Gamma _0},
\ee
where
\be \label{alpha}
\Couleur{\alpha ^0 \equiv \gamma ^0/g^{00}}, \qquad \Couleur{\alpha ^j \equiv \gamma ^0\gamma ^j/g^{00}}.
\ee
The spin connection matrices with an orthonormal tetrad field $\Couleur{(u_\alpha )}$ are given by: 
\be
\Couleur{\Gamma ^\sharp_\epsilon  =\frac{1}{8}\gamma _{\alpha \beta \epsilon }\left [ \gamma ^{\sharp \alpha },\gamma ^{ \sharp \beta }\right]}.\qquad (\Couleur{\gamma ^{\sharp \alpha }=}\mathrm{``flat"\ Dirac\  matrices})
\ee
(See e.g. Ref. \cite{Ryder2008}.) Therefore, with the natural basis ($\Couleur{\partial _\mu =b^\alpha _{\ \,\mu } u_\alpha )}$, they become:
\be
\Couleur{\Gamma _\mu =b^\alpha _{\ \,\mu } \Gamma ^\sharp_\alpha}.
\ee
Using the foregoing expressions, it is straightforward to compute $\Couleur{\mathrm{H}}$ in the rotating frame $\Couleur{\mathrm{F}}$ with the adapted rotating tetrad (\ref{rotating tetrad}). We find that the spin connection matrices $\Couleur{\Gamma _\mu}$ do involve spin operators made with the Pauli matrices $\Couleur{\sigma ^j}$. In particular, we have for $\Couleur{V\equiv \omega \rho \ll c}$: 
\be
\Couleur{\Gamma _0=-\frac{i}{2}\frac{\omega }{c}\Sigma ^3 \left [1+O\left(\frac{V}{c}\right)\right], \qquad \Sigma ^j\equiv \begin{pmatrix} 
\sigma ^j & 0\\
0  &  \sigma ^j
\end{pmatrix}},
\ee
for which $\Couleur{-i\hbar c\Gamma _0}$ is the usual {\it ``spin-rotation coupling" term} \cite{HehlNi1990,Ryder2008} in $\Couleur{\mathrm{H}}$. We find that also the $\Couleur{\Gamma _j}$ matrices ($\Couleur{j=1,2,3}$) contain spin operators. This is likely to come from the fact that the adapted rotating tetrad involves projecting the natural tetrad of the rotating coordinates.\\

On the other hand, the Minkowski metric has obviously the form (\ref{space-isotropic diagonal}). Thus, let us now evaluate again the Hamiltonian $\Couleur{\mathrm{H}}$ in the rotating frame, but this time from the ``diagonal tetrad" prescription (\ref{diag tetrad}) in the Cartesian chart $\Couleur{\chi '}$, which corresponds with choosing the Minkowski tetrad, i.e., the natural basis \Couleur{$(\partial'_\mu )$} of the chart $\Couleur{\chi '}$. Defining the $\Couleur{\gamma^\mu  }$ matrices in the rotating chart $\Couleur{\chi }$ from using that tetrad, gives after a simple calculation:
\be \label{Hamilton-restricted-gauge}
\Couleur{\mathrm{H} =  \mathrm{H}'-i\hbar \omega (y\partial _x-x\partial _y)=\mathrm{H}'-\Mat{\omega .}{\bf L}},
\ee
with $\Couleur{\mathrm{H}'}$ the special-relativistic Dirac Hamiltonian in the inertial frame $\Couleur{\mathrm{F}'}$, and $\Couleur{{\bf L}\equiv {\bf r}\wedge (-i\hbar \nabla )}$ the angular momentum operator. (We note that the same $\Couleur{\mathrm{H}}$ applies, whether \hyperlink{Spin depends on gamma}{DFW} or \hyperlink{QRD-0}{QRD--0} is chosen. The spin connection matrices are zero.) Thus, there is no spin-rotation coupling with the ``constant gauge transformations" solution of the non-uniqueness problem.

%%%%%%%%%%%%%%%%%%%%%%%%%%%%%%%%%%%%%%%%%%%%%%%%%%%%%%%%%%%%%%%%%%%%%%%%%%%%%%%%

%%%%%%%%%%%%%%%%%%%%%%%%%%%%%%%%%%%%%%%%%%%%%%%%%%%%%%%%%%%%%%%%%%%%%%%%%%%%%%%%
  \section{Conclusion}
%%%%%%%%%%%%%%%%%%%%%%%%%%%%%%%%%%%%%%%%%%%%%%%%%%%%%%%%%%%%%%%%%%%%%%%%%%%%%%%%{\bf 

The (generally-)covariant Dirac theory leads to non-unique Hamiltonian and energy operators in any given coordinate system. This is due to the gauge freedom that exists in the choice of the $\Couleur{\gamma^\mu  }$ matrices, or equivalently in the choice of the tetrad field \Couleur{$(u_\alpha )$}. This non-uniqueness is there despite the fact that, by construction, the standard covariant Dirac {\it equation} is independent of the choice of the tetrad field (in a topologically-simple spacetime).\\

A rather ``conservative" way of restricting the gauge freedom so as to get unique Hamiltonian and energy operators, is to fix the time-like vector $\Couleur{u_0}$ of the tetrad, and then to fix the rotation rate tensor field \Couleur{$\Mat{\Xi}$} of the spatial triad $\Couleur{(u_p)}$. This applies only to a given reference frame. Also, it is uneasy to implement. In the archetypical case of the uniformly rotating frame in a Minkowski spacetime, this way leads to the presence of a spin-rotation coupling term in the Hamiltonian, provided the rotation rate field \Couleur{$\Mat{\Xi}$} is fixed to be the rotation rate tensor field \Couleur{$\Mat{\Omega }$} of the reference frame itself. \\

A more ``radical" solution of the non-uniqueness problem is to arrange that the same gauge freedom applies as in special relativity---constant gauge transformations. This needs that the metric can be put in the diagonal space-isotropic form (\ref{space-isotropic diagonal}). (See Ref. \cite{A48} for a justification of the physical relevance of this metric.) This solution then applies independently of the reference frame. Moreover, it is easy to implement. However, in a uniformly rotating frame in a Minkowski spacetime, this solution leads to the absence of any spin-rotation coupling term in the Hamiltonian.

%%%%%%%%%%%%%%%%%%%%%%%%%%%%%%%%%%%%%%%%%%%%%%%%%%%%%%%%%%%%%%%%%%%%%%%%%%%

%%%%%%%%%%%%%%%%%%%%%%%%%%%%%%%%%%%%%%%%%%%%%%%%%%%%%%%%%%%%%%%%%%%%%%%%%%%
\end{document}